\documentclass[aps,prl,twocolumn,showpacs]{revtex4}

\usepackage{epsfig}
\usepackage{graphicx}

\begin{document}

\DeclareGraphicsExtensions{.eps,.EPS}

\title{Averaging out magnetic forces with fast rf-sweeps in an optical trap for metastable chromium atoms.}
\author{Q. Beaufils, R. Chicireanu, A. Pouderous, W. de Souza Melo$^{\dag }$, B. Laburthe-Tolra, E. Mar\'echal, L. Vernac, J. C. Keller, and O. Gorceix}
\affiliation{Laboratoire de Physique des Lasers, CNRS UMR 7538, Universit\'e Paris 13,
99 Avenue J.-B. Cl\'ement, 93430 Villetaneuse, France}
\affiliation{$^{\dag }$ Departamento de F\'isica, Instituto de Ci\^encias Exatas, Universidade Federal de Juiz de Fora, Juiz de Fora - MG - Brasil
 }

\begin{abstract}
We introduce a novel type of time-averaged trap, in which the \textit{internal} state of the atoms is rapidly modulated to modify magnetic trapping potentials. In our experiment, fast radiofrequency (rf) linear sweeps flip the spin of atoms at a fast rate, which averages out magnetic forces. We use this procedure to optimize the accumulation of metastable chomium atoms into an optical dipole trap from a magneto-optical trap. The potential experienced by the metastable atoms is identical to the bare optical dipole potential, so that this procedure allows for trapping all magnetic sublevels, hence increasing by up to 80 percent the final number of accumulated atoms. 
\end{abstract}


\date{\today}

\maketitle

In recent years, the use of radio-frequency (rf) fields in combination with magnetic static fields was suggested to engineer cold atoms traps with versatile shapes. The original idea \cite{ZG} is to use the avoided crossing due to the coupling of different magnetic states resonant at a given magnetic field for a given rf frequency to adiabatically deform a magnetic trap and enhance the trapping in one direction. O. Zobay and B. M. Garraway originally suggested to prepare a 2D Bose-Einstein condensate; indeed Y. Colombe et al. confined ultra cold atoms in rf-induced 2D trapping potentials \cite{Colombe04}. This scheme is promising, as Bose-Einstein Condensation in 2D is different than in 3D: in 2D, superfluidity emerges through the Berezinskii-Kosterlitz-Thouless phase transition \cite{BKT}. In the same spirit, rf adiabatic potentials were used to split quasi-1D BECs in a double-well \cite{schmidtketterle}. Rf-induced adiabatic potentials could also be used to create ring traps for ultra-cold atoms \cite{Morizot06}, or to build arrays of microtraps with lattice constants well below 1 $\mu$m \cite{Courteille}. Thus, rf traps offer a relatively easy means to deform magnetic traps in a time-dependent way, with large versatility and flexibility (see also \cite{Lesa}).

In this Letter, we introduce a new way of tailoring rf fields to modify magnetic trapping: we produce fast rf-sweeps to time-average magnetic forces, by forcing the atoms to oscillate between different magnetic sub-levels which experience different magnetic forces. In contrast to other time-averaged traps, such as TOP traps \cite{top}, or the Paul trap \cite{paul}, where the spring constant of a potential is modulated, here, it is the internal state of the atoms which is modulated in time. In our experiment, chromium atoms are trapped in a Magneto-Optical Trap (MOT). Atoms from the MOT are optically pumped into metastable dark states $D$, which are trapped by the combination of an optical trap (OT), and a magnetic trap due to the magnetic quadrupole field used to produce the MOT \cite{Chicireanu07-3}. Along the longitudinal direction of the OT, the magnetic forces dominate the optical dipole forces, which, combined to the radial confinement of the OT, create a strongly confining potential in all directions. Because metastable chomium atoms have a large inelastic loss parameter (\cite{schmidt},\cite{Chicireanu07-2}), this strong 3D confinement limits the number of atoms in the mixed trap \cite{Chicireanu07-3}.

\begin{figure}
\centering
\includegraphics[width= 2.8 in]{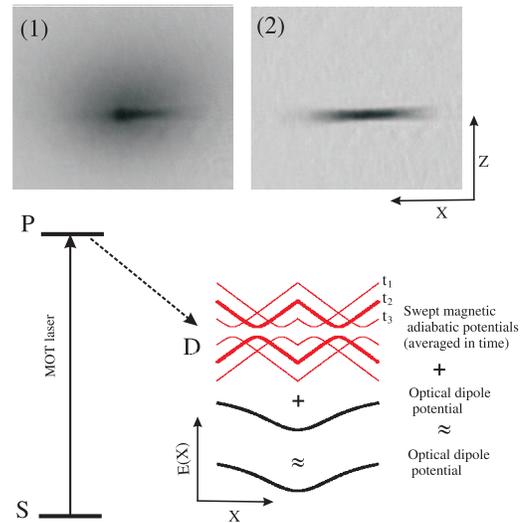}
\caption{\setlength{\baselineskip}{6pt} {\protect\scriptsize
Absorption pictures of the atoms in the mixed trap: (1) without rf; (2) with rf. The sketch describes the principle of this experiment. MOT atoms decay into $D$ states from the excited state. Two time-dependent adiabatic potentials of the $D$ states are qualitatively represented for three different times $t_1$, $t_2$, $t_3$, and two magnetic numbers $m_J$ and $-m_J$}} \label{Fig1}
\end{figure}

With the use of linear rf-sweeps described in this Letter, magnetic forces are averaged to zero, and the net trapping potential is identical to the one of a pure optical trap: atoms accumulate in an optical trap free of magnetic forces, which considerably reduces the longitudinal confinement. This new strategy has various advantages: (i) all magnetic sublevels are trapped; (ii) the volume of the trap is larger than the volume of the mixed magnetic plus dipole trap, which reduces the peak atom density for the same number of atoms, and therefore the effect of inelastic collisions; (iii) as the atoms directly accumulate in the OT potential, the question of mode-matching the magnetic trap and the OT, which is usually difficult, disappears. (iv) In addition, metastable atoms are completely decoupled from the MOT, as they experience neither the MOT light, nor the MOT magnetic gradient. In fact, as we will show below, we also can neglect collisions between metastable atoms and atoms in the MOT on the loading timescale.

Here, we demonstrate that the use of fast rf-sweeps to time average magnetic forces to zero allows us to load up to 80 percent more atoms in the dipole trap than without these rf sweeps. We observe up to 2 $10^6$ chomium atoms loaded in about 100 ms, which should be a good starting point to produce a chromium BEC by forced evaporation. This experimental result is therefore an important step, given the interest of chomium atoms for the study of dipole-dipole interactions in quantum degenerate Bose \cite{pfau} and Fermi gases.

Our experimental set-up was already described in \cite{Chicireanu06-1}. We collect up to 4.10$^6$ chromium atoms in a magneto-optical trap, at a temperature of 120 $\mu$K. Cold atoms slowly decay into metastable dark states. These remain trapped in the magnetic gradient used for the MOT, provided their magnetic number $m$ is positive. This allows for the accumulation of large numbers of trapped chomium atoms, despite a very large light-assisted collision rate in the MOT. As studied in \cite{Chicireanu07-3}, when a 1075 nm, 35W, retro-reflected fiber laser beam is focussed on the MOT (with a waist of 42 $\mu$m at the atoms' position), up to 1.2 10$^6$ metastable atoms with $m>0$ accumulate in a mixed magnetic plus far-detuned dipole optical trap when the OT is on during the cooling stage. However, inelastic and Majorana losses in the metastable states limit the total number of atoms trapped, as well as the peak density. Very few atoms in the $S$ states are trapped in the OT while the MOT light is on, presumably because light-assisted collisions noticeably limit the lifetime of dense clouds of ground-state atoms. 

In Fig \ref{Fig1}, we show two absorption images taken just after the MOT lasers are switched off and all metastable atoms are repumped to the electronic ground-state. Figure \ref{Fig1} (a) clearly features the presence of the dipole trap (where the density is highest), surrounded by a halo of magnetically trapped atoms. Figure \ref{Fig1}(b) is an absorption image taken when fast rf-sweeps are continuously applied to the atoms while they accumulate in the trap. The rf is sent to the atoms using a 8 turn, 8 cm diameter coil located 4.5 cm away from the atoms. Its symmetry axis is perpendicular to the long axis of the dipole trap. In presence of rf, no atom remains in the magnetic trap. However, the number of atoms trapped in the OT is actually increased. The density profiles show that the rf produces an increase of the trap volume due to a larger extension along the longitudinal axis of the OT, allowing for a larger number of atoms to be loaded for a given peak central density. Temperature measurements with and without rf sweeps give the same temperature (100 $\mu$K $\pm$ 10 $\mu$K), indicating that rf sweeps do not produce noticeable heating on the accumulation timescale. We also note that the presence of rf does not modify the MOT properties. Indeed, the rf Rabi frequency is always smaller than the optical pumping rates in the MOT. In addition, as the rf frequency is swept, at any given moment, only a small fraction of the atoms are resonant with the rf.

\begin{figure}
\centering
\includegraphics[width= 2.8 in]{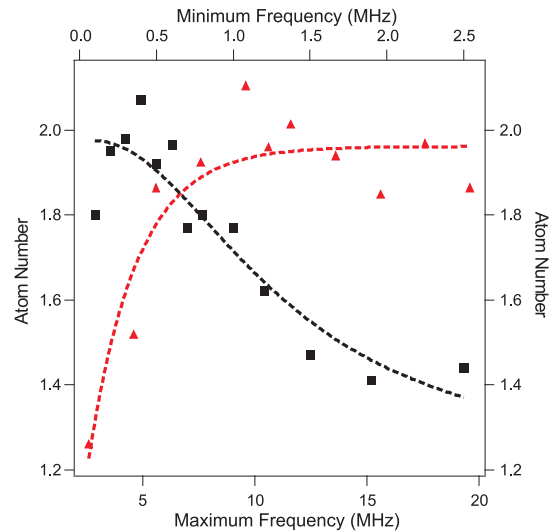}
\caption{\setlength{\baselineskip}{6pt} {\protect\scriptsize
Steady-state atom number in the optical trap as a function of minimum (squares) and maximum (triangles) rf frequency in the sweep (for a mean rf power of 80 W, and a sweep rate of the rf frequency of 10 kHz). Lines are guides for the eye.}} \label{Figmaxmin}
\end{figure}

We studied how the atoms are trapped in the optical trap for different rf sweeps, varying the total power, the maximum and minimum frequencies in the sweep, as well as the sweep rate. We also varied the sweep shape, and found that linear ramps are optimal (either increasing or decreasing seesaw ramps) to increase the number of trapped atoms. We first found that (for a large enough rf power, as discussed below) the most efficient frequency span corresponds to sweeping between $\nu_{min} = $500 kHz or below, to $\nu_{max} = $7 MHz or above (see Fig (\ref{Figmaxmin})). When $\nu_{min}$ is below 500 kHz, the atomic signal does not change much. Such low frequencies are resonant with atoms close to the center of the trap. For these locations, the magnetic field set by the MOT gradient is smaller than the rf magnetic field amplitude, so that one cannot use the rotating wave approximation to describe the system. For the atoms close to $B=0$, a rf frequency of 500 kHz is blue detuned compared to their Zeeman structure. In this regime, it is shown in \cite{haroche} that atoms are strongly coupled to the rf, provided that the Rabi frequency is at least on the order of the rf frequency. In practice, $\nu_{min}$ is set to 500 kHz for data shown in the rest of this paper. 

On the other hand, the maximal rf frequency in the sweep $\nu_{max}$ is set to be resonant with atoms at the edge of the cloud in the dipole trap, so that the rf sequentially couples to all atoms in the dipole trap as the frequency is scanned. $\nu_{max}$ is qualitatively given by 

\begin{equation}
h \nu_{max} \approx g_J \mu_B B' z_R
\label{eqmax}
\end{equation}
where $g_J$ is the Land\'e factor, $\mu_B$ the Bohr magneton, $B'$ the magnetic field gradient (9 G/cm), and $z_R$ is the 2.5 mm Rayleigh length of the trapping laser. This criterion gives $\nu_{max} > 5$ MHz, in qualitative agreement with our observation that the signal saturates for $\nu_{max} > 7$ MHz.

\begin{figure}
\centering
\includegraphics[width= 2.8 in]{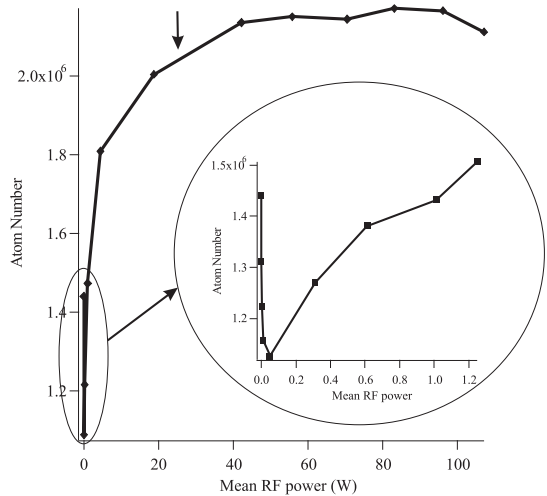}
\caption{\setlength{\baselineskip}{6pt} {\protect\scriptsize
Steady-state atom number in the OT as a function of the mean rf power, for rf frequency sweeps between 500 kHz and 7 MHz, at a rate of 10 kHz; insert: same, for low rf power, and similar experimental conditions. Arrow : typical RF power value above which the adiabatic criterion (\ref{eqLZ}) is fulfilled}} \label{Fig2}
\end{figure}

We analyzed how the final number of atoms depends on the rf power, as shown in Figure \ref{Fig2}. The number of atoms in the OT first decreases with increasing rf power. For such low rf power, the rf sweeps are not adiabatic, but the rf field is strong enough to change the magnetic sublevel of the atoms during the accumulation timescale: positive $m$ states couple to negative $m$ states, which are expelled. When the rf power is larger, however, the rf sweeps are fully adiabatic, and the rate at which the spin states flip, set by the sweep rate, is high enough for the magnetic force to be averaged to zero. The net potential experienced by the atoms is that of a pure OT.

Simple arguments give the order of magnitude of the rf Rabi angular frequency $\Omega= g_J \mu_B B_{rf}/\hbar$ needed to spin flip all atoms in a single sweep (where $B_{rf}$ is the rf magnetic field amplitude). In simple terms, the rf field will efficiently couple atoms in a region $\Delta x$ such that $g_J \mu_B B' \Delta x \approx \hbar \Omega$. If the rf sweep amplitude is $\Delta \nu$, the rf sweep couples to all atoms between $x=0$ and $x=L$ such that $g_J \mu_B B'L \approx h \Delta \nu$. As a consequence, if the sweep duration is $t_S$, atoms in the region $\Delta x$ feel the rf for a duration $\frac{\Delta x}{L} t_S$, and the condition for all atoms to be efficiently flipped is $\frac{\Delta x}{L} t_S > 2 \pi / \Omega$, which reads:

\begin{equation}
\left(\frac{\Omega}{2 \pi}\right)^ 2 \geq \alpha \Delta \nu / t_S
\label{eqLZ}
\end{equation}

where $\alpha$ is a numerical factor. One therefore recovers the usual Landau Zener criterion for adiabatic crossing. We solved the time dependent Schrodinger equation for the $^5 D _4$ state, and find that rf sweeps are fully adiabatic when Eq. (\ref{eqLZ}) is fulfilled with $\alpha = 4$. In our experiment, $\Omega$ depends on the rf frequency, as the rf coil impedance changes with the rf frequency. It is therefore difficult to directly compare this criterion to the data plotted in Figure \ref{Fig2}. As an indication, we nevertheless show by an arrow in Figure \ref{Fig2} the value of the rf amplitude above which we expect the adiabatic criterion to be fulfilled, at the mean rf frequency of 3 MHz. The rf power sent to the atoms is calibrated by measuring the induction voltage in a small loop located at the center of the rf coil. The saturation in the number of trapped atoms is clearly achieved when inequality (\ref{eqLZ}) is verified. Then, the spin-flip rate is equal to the sweep rate \cite{notequad}. 

\begin{figure}
\centering
\includegraphics[width= 2.8 in]{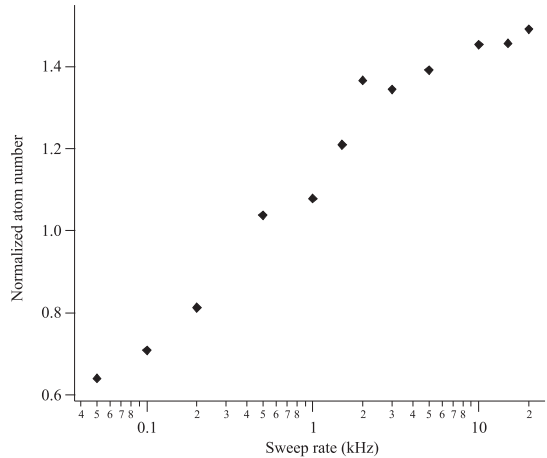}
\caption{\setlength{\baselineskip}{6pt} {\protect\scriptsize
Steady-state atom number in the optical trap as a function of the rf frequency sweep rate, for an average rf power of 100 W, with $\nu_{min}$ = 500 kHz, and $\nu_{max}$ = 7.5 MHz. Atom numbers are normalized by reference to the OT population with no rf.}} \label{Fig3}
\end{figure}

We now turn to the question of the minimum necessary sweep rate to efficiently average magnetic forces to zero. We plot in Figure \ref{Fig3} the steady state atom number after accumulation, as a function of the sweep rate. If the sweep rate is too small, there are less atoms loaded in the trap than without rf. For large enough rf power, the atom number increases with the rf sweep rate and starts saturating for rates larger than 1 kHz (mind the logarithmic scale of Fig. (\ref{Fig3})). 

Qualitatively, we expect that magnetic forces are averaged to zero if the spin flip rate (given by the sweep rate $1/t_S$ provided the rf coupling $\Omega$ is large enough) is large compared to the oscillation frequency of a positive magnetic sublevel $m$ in the quadrupole trap. It is difficult to give an estimate of the typical oscillation time of atoms in the magnetic trap, as this trap is not harmonic. In addition, different $m$ states experience different trapping strengthes. As a qualitative indication, we note that the typical $1/e$ radius of the pure magnetic trap is 500 $\mu$m, at a temperature of 100 $\mu$K. Assuming thermal equilibrium, this shows that the average magnetic sublevel $m$ is about 2, and the average oscillation time in the magnetic gradient about 10 ms. Therefore, one can expect the sweep averaging to be efficient for $1/t_S >$ 1 kHz, in good agreement with our observations (see Fig. \ref{Fig3}).

To summarize, atoms will efficiently be stored in the rf-swept trap if: (i) \textit{all} atoms are spin-flipped at a fast rate; (ii) the spin flip rate is higher than the oscillation rate of atoms in the pure magnetic trap. Provided these conditions are fulfilled, linear sweeps insure that the effect of magnetic forces on atoms is averaged out. It is possible that other rf sources, such as a white-noise rf source, of a rf frequency comb, may be as efficient, or even more efficient, to cancel magnetic forces. However, power requirements for both these solutions may be much more stringent than for the use of linear sweeps, because the total rf power is shared between all the rf modes, and the Landau-Zener adiabatic criterion for the avoided crossings may be harder to fulfill. In our experiment, less atoms were accumulated using rf combs than using linear sweeps. 

We now turn to the loading dynamics of the rf-swept magnetic plus dipole trap. We have performed experiments where we have accumulated atoms for various durations. The 1/e loading time is 120 ms, for a steady state number of atoms of 2.10$^6$ atoms, corresponding to a loading rate of 1.7.10$^7$ atoms/s. This large value is close to the loading rate of our magnetic trap, reported in \cite{Chicireanu07-2} for similar experimental conditions, with no rf and no OT. The situation is nevertheless quite different here, first because of the small overlap between the MOT and the dipole beam (whose transverse size is smaller than the MOT radius), second because the present loading scheme traps all $m$ states. 

To understand what sets the loading time in our experiment, we performed life time measurements. We found that the 1/e lifetime of the cloud in the rf-swept-magnetic plus dipole trap is close to the 1/e loading time. This rules out inelastic collissions with the MOT as the limiting factor (in contrast with \cite{Chicireanu07-2}). This implies that atoms in the metastable states are now completely decoupled from the MOT, in the sense that they feel neither the MOT atoms nor the MOT trapping forces. 

At short time, the decay of atoms in the rf-swept-magnetic plus dipole trap is close to the decay of atoms in the pure OT. This indicates that the main loss process in our system is not related to rapid spin flipping of the atoms. We do observe a difference for larger times, though, which may indicate some heating due to the rf sweeps. The careful study of this effect is beyond the scope of this paper, and will be studied in more detail in the near future. Using  our previous measurements of the elastic cross-section and inelastic loss parameter for metastable chromium atoms \cite{Chicireanu07-2}, we infer that both free evaporation and inelastic losses limit the accumulation time. 

In conclusion, we have demonstrated a new technique to time-average magnetic forces. Our technique makes use of linear rf sweeps to resonantly spin flip all atoms in a magnetic field gradient, at a high rate. This procedure increases the volume of the trap in which metastable chromium atoms are accumulated from the MOT. In addition, it has the advantage to trap all magnetic sublevels in the same trap, despite the presence of a magnetic field gradient.

Acknowledgements: LPL is Unit\'e Mixte (UMR 7538) of CNRS and of Universit\'e Paris Nord. We acknowledge financial support from Conseil R\'{e}gional d'Ile-de-France (Contrat Sesame), Minist\`{e}re de l'Enseignement Sup\'{e}rieur et de la Recherche, European Union (FEDER -Objectif 2), and IFRAF (Institut Francilen de Recherche sur les Atomes Froids).


\begin{thebibliography}{99}

\bibitem{ZG} O. Zobay and B. M. Garraway, Phys. Rev. Lett. \textbf{86}, 1195 (2001)

\bibitem{Colombe04} Y. Colombe et al.,  Europhys. Lett. \textbf{67}, 593 (2004). See also M. White et al., Phys. Rev. A \textbf{74}, 023616 (2006).

\bibitem{BKT} V. L. Berezinskii, Sov. Phys. JETP \textbf{32}, 493 (1971); \textbf{34}, 610 (1972). J. M. Kosterlitz and D. J. Thouless, J. Phys. C., \textbf{6}, 1181 (1973). Z. Hadzibabic et al. Nature (London) \textbf{441}, 1118 (2006).

\bibitem{schmidtketterle} S. Hofferberth et al., Nature Physics \textbf{2} , 710 (2006).

\bibitem{Morizot06} O. Morizot et al., Phys. Rev. A \textbf{74}, 023617 (2006).

\bibitem{Courteille} Ph. W. Courteille et al., J. Phys. B \textbf{39}, 1055 (2006).

\bibitem{Lesa} I. Lesanovsky and W. von Klitzing, Phys. Rev. Lett. \textbf{99}, 083001 (2007). 

\bibitem{top} W. Petrich et al., Phys. Rev. Lett., \textbf{74}, 3352 (1995)

\bibitem{paul} W. Paul, Rev. Mod. Phys., \textbf{62}, 531 (1990)

\bibitem{Chicireanu07-3} R. Chicireanu et al., Eur. Phys. J. D., \textbf{45}, 189 (2007). 

\bibitem{schmidt} P. O. Schmidt et al., J. Opt. B.: Quantum Semiclassical Opt., \textbf{5}, S170 (2003).

\bibitem{Chicireanu07-2} R. Chicireanu et al., Phys. Rev. A \textbf{76}, 023406 (2007). 

\bibitem{pfau} T. Lahaye et al., Nature \textbf{448}, 672 (2007).

\bibitem{Chicireanu06-1} R. Chicireanu et al., Phys. Rev. A, \textbf{73}, 053406 (2006).

\bibitem{haroche} S. Haroche et al., Phys. Rev. Lett., \textbf{24}, 861 (1970) and references therein.

\bibitem{notequad} A complication arises from the tensor light-shift of the metastable states in the dipole trap. Different $m$ states experience slightly different dipole potentials \cite{Chicireanu07-3}. When the rf Rabi frequency is higher than the differential light shift between adjacent $m$ states, this effect can be neglected, which is indeed the case in our experiment for large enough rf power. 



\end{thebibliography}
\end{document}